\documentclass[a4paper]{article}
\usepackage{graphicx}
\usepackage{amssymb}
\usepackage{amsfonts}

\title{Quantum Newtonian cosmology revisited}

\author{Laure Gouba \\
Abdus Salam International Centre for Theoretical Physics,                       ICTP \\ Strada Costiera, 11, I-34151 Trieste Italy\\
Email: lgouba@ictp.it}

\begin{document}

\maketitle

\begin{abstract}
We formulate the Lagrangian of the Newtonian cosmology where the cosmological constant is also introduced. Following the affine quantization procedure,  the Hamiltonian operator is derived. The wave functions of the Newtonian universe and the corresponding  eigenvalues for the case of matter dominated by a negative cosmological constant are given.
\end{abstract}

\section{Introduction}

During the past three years, I have been teaching the course of concepts in physics and problems solving in physics at the
African Institute for Mathematical Sciences (AIMS) Centres at Mbour/Senegal (2018) and Accra/Ghana (2019, 2020). The course is basically about Newtonian physics. Like the students, I found the basic astronomy concepts fascinating. 

Defining the universe as all of space and time and their contents, including planets, stars, galaxies, and all other forms of matter and energy, some natural questions arise about the beginning, the existence of the universe, its present constituents and how will be the life in the future. The field of cosmology is the best candidate to respond to those questions as cosmology seeks to answer to these questions that are the oldest of the mankind. The study of the cosmos is as old as humanity and has always been fascinating.
Physical cosmology is the scientific study of the universe as a whole based on the laws of physics, to which Einstein's equations are very likely to give a correct description. General relativity describes the observed universe very well.
However, it is a complicated theory and often it is difficult to find solutions. The Newtonian theory, on the other hand, is in many ways far simpler and is very often able to satisfactorily model the universe. 

The Newtonian cosmology began with the work of McCrea and Milne in 1934 in which simple isotropic and homogeneous universe were studied \cite{milne, crea}. The two assumptions fit well with the observable universe, namely the ex\-pan\-ding universe free of rotation and shear.
There are two approaches to Newtonian cosmology: the first one
who deals with potential function worked out by McCrea and Milne \cite{ milne, crea}, G\"odel \cite{godel}, Herkmann and Sch\"ucking \cite{herk1,herk2}, Raychaudhury \cite{ray}, Zeldovich \cite{zeldo}.  
The second approach who uses gravitational force
is given by Narlikar \cite{narli} . 
Summary of Newton cosmology is given by Bondi \cite{bondi}, North \cite{north}, Ellis \cite{ellis}.

The application of quantum theory to the description of the universe as a whole is called quantum cosmology. Quantum Newtonian cosmology was proposed by Freedman et al.\cite{freedman}. It is possible to construct a wave function for Newtonian cosmology in the framework of non-relativistic quantum mechanics \cite{jnorb, juan, bras2, bras3}. We are interested in quantum cosmology and a good way to start is to revisit quantum Newtonian cosmology in the sense of \cite{bras2}.

Our work is presented as follows:
In section (\ref{sec2}) we present the classical model, where  we start by recalling the cosmological principle and then formulate the Lagrangian like in \cite{bras1}. In section (\ref{sec3}), we perform the affine quantization procedures and formulate the quantum Hamiltonian operator. In section (\ref{sec4}) the wave functions of the Newtonian universe and the energy levels are given. The  section (\ref{sec5}) is about some concluding remarks.  An appendix about Biconfluent Heun equation and its polynomial form of solution is added after the concluding remarks. 

\section{The classical model}\label{sec2}

\subsection{Cosmological principle}

In modern physics, cosmology begins with the application of Einstein's theory of gravity, or general relativity, to the universe. This is a difficult task and would probably not be possible without a basic assumption about the universe called the cosmological principle that says that `` On large (but not too large) scales, the universe is homogeneous and isotropic ". The statement of isotropy means that the universe is the same in all directions (the universe looks the same whether you are looking directly outward from the North Pole or the South Pole. The Homogeneity means that the universe is the same at all points. 
The statements of homogeneity and isotropy are distinct but closely related: for example a universe which is isotropic will be homogeneous while a universe that is homogeneous may not be isotropic. A universe which is only isotropic around one point is not homogeneous. Without the cosmological principle, much of our presumed understanding of the workings of the universe would be invalid.

\subsection{Lagrangian formulation}

We consider a cosmological model based on Newtonian dynamics. It uses an approach worked out by Milne and McCrea in which it has no sense to associate the gravitational phenomena to the effects of spacetime curvature \cite{milne, crea}. In this context, it was shown that the universe behavior could be understood on the basis of classical physics, which does not use the mathematical complexity in the study of the universe as in the Einsteinian cosmology.
The simplest model of the universe one can think of is that of a universe filled with dust (non-relativistic pressureless matter, $p=0$). We can think of this dust as a collection of point particles, and on the cosmological scales of approximation 10-1000Mpc. These point particles are a sufficiently good approximation for galaxies or even galaxy clusters.
Consider the evolution of a small spherical portion of the universe, where Newton's theory applies: then the behavior of this portion will reflect the evolution of the universe.

The cosmological principle states that there is no prefered place or direction in the universe on large scales, so we can pick any coordinate system with respect to which we can measure the positions and velocities of these test particles.

Let us consider a galaxy that we consider as a particle of mass $m$ for our Newtonian system located at a radius $a(t)$ from an arbitrary defined origin, where $a(t)$ is  a distance we 
refer to as the scale factor. This scale factor acts to simply scale up or scale down our Newtonian universe. We assume that the motion about the origin must be spherically symmetric. 
The force of gravity on the particle (galaxy) at distance $a(t)$ coming from the mass of the homogeneous universe inside the sphere of radius $a(t)$ is the same as if all the mass were at the center of the sphere. There is no force arising from the region outside the sphere.

In a fixed rectangular coordinates (comoving), the kinetic energy for the motion of the particle (galaxy ) is
 given by 
 \begin{equation}
 T = \frac{1}{2}m\dot{a}^2\,.
 \end{equation}
 We assume that the particle (galaxy) moves in a conservative force field so the potential energy  is given by 
 \begin{equation}
 U = - \frac{GMm}{a}\,.
 \end{equation}
In both Newtonian and relativistic cosmology, the universe is unstable to gravitational collapse. Both Newton and Einstein believed the universe is static. In order to obtain this, Einstein introduced a repulsive gravitational force
\begin{equation}
F_\Lambda = \frac{\Lambda}{3} m a,
\end{equation}
where the constant $\Lambda$ is called the cosmological constant. When $\Lambda > 0$, this force is pointed radially outward, repulsive relative to the point origin. If $\Lambda < 0$, the force is attractive relative to the point origin. We can also introduce in the Newtonian approach to cosmology a term containing the cosmological constant, associated with a kind of cosmological force. We assume then that there exists a global cosmological force that affects the particle (galaxy). We take this force to be the repulsive gravitational force that yields the potential 
\begin{equation}
U_\Lambda  = -\int_0^a F_\Lambda da'  = -\frac{1}{6}\Lambda m a^2.
\end{equation}
In a fixed rectangular coordinates (comoving) the Lagrangian of the system is 
\begin{equation}\label{clagrang}
L (a, \dot{a}) = \frac{1}{2}m\dot{a}^2 + \frac{GMm}{a} +
 \frac{1}{6}\Lambda m a^2.
\end{equation}
The Lagrangian depends only on the scale factor and its derivative, and is independent of the time because the system is under the action of a uniform force field. 

\section{Affine quantization and Hamiltonian operator}\label{sec3}

The constant quantity of the motion is the classical Hamiltonian of the system defined by 
\begin{equation}\label{chamil1}
H = \frac{\partial L}{\partial \dot{a}}\dot{a} - L\,.
\end{equation}
Inserting equation (\ref{clagrang}) into the equation (\ref{chamil1}) we got the Hamiltonian at the classical level
\begin{equation}\label{chamil2}
H(a, p_a) = \frac{1}{2m}p_a^2  - \frac{G\;M\; m}{a} - \frac{1}{6}\Lambda m a^2\,,
\end{equation}
where $p_a$ is considered as the corresponding momentum
\begin{equation}\label{mom}
p_a = \frac{\partial L}{\partial \dot{a}} = m\dot{a}\,.
\end{equation}

We consider a fixed rectangular comoving coordinate in which we can define the phase space as $(a, p_a)$ with $p_a$ in (\ref{mom})
being the linear momentum. 
The associate Hamiltonian in terms of the phase space coordinates is given by (\ref{chamil2}).
Since the variable $a$ is positive, we consider the positive axis. In that sense we choose to perform the affine quantization procedure  due to J. R. Klauder \cite{klauder1, klauder2, klauder3} that received some recent  applications \cite{fanuel,bergeron, zonetti, almeida, laure1, laure2, riccardo}. Here we choose to pay a careful attention to the quantization procedures.

The non-vanishing Poisson bracket of the canonical coordinates is given by 
\begin{equation}\label{npb}
\left\lbrace a,\; p_a\right\rbrace = 1\,.
\end{equation}
Multiplying the equation (\ref{npb}) by $a$, we have 
\begin{equation}
a \left\lbrace a,\;p_a\right\rbrace = a\,,
\end{equation}
and setting $d_a = a p_a $, we have 
\begin{equation}
\left\lbrace a,\;d_a\right\rbrace = a,
\end{equation}
 and
\begin{equation}
H (a, d_a)= \frac{1}{2}d_a\; a^{-2} d_a - \frac{G\,M\,m}{a} - \frac{\Lambda}{6}\; m\; a^2\,.
\end{equation}
The variables $a,\; d_a$ are not canonical variables but 
form a Lie algebra and then worthy to be considered as a new 
pair of classical variables. The canonical quantization involves 
$\hat a,\; \hat p_a$ which are self adjoint operators that satisfies the commutation relations 
\begin{equation}
\left[ \hat{a}, \hat{p}_a\right] = i\hbar,
\end{equation}
and from the canonical quantization, it follows that 
\begin{equation}
\left[ \hat a, \hat{d}_a \right] = i\hbar \hat a,
\end{equation}
$\hat d$ is the dilation operator and the operators $\hat a$ and $\hat{d}_a$ are realized as follows
\begin{equation}
\hat a \psi (a) = a\psi(a);\quad 
\hat{d}_a \psi (a) = -i\hbar\left( a\frac{d}{d a} +\frac{1}{2}\right)\psi(a)\,,
\end{equation}
where $\psi(a)$ is the wave function of the Newtonian universe. The  Hamiltonian operator for a particle (galaxy) moving in the Newtonian universe is then given by 
\begin{equation}
\hat{H} = -\frac{\hbar^2}{2m}\frac{d^2}{da^2} + 
\frac{\hbar^2}{2m}\frac{3}{4}\frac{1}{a^2} -\frac{G\;M\;m}{a} 
- \frac{\Lambda}{6}ma^2\,.
\end{equation}

\section{The Wave functions of the Newtonian universe and the energy levels}\label{sec4}

Once we have the Hamiltonian operator of the system, it is possible to find the wave functions of the Newtonian universe \cite{juan, bras2, bras3}, by solving the time-independent Schr\"odinger equation
\begin{equation}\label{tise}
\hat H \psi(a) = E \psi(a)\,,
\end{equation}
where $\psi(a)$ corresponds to the eigenvalues $E$, with 
$\Psi(a,t) = \psi(a)e^{-iEt/\hbar}$ being the general wave function solution of the time-dependent Schr\"odinger equation.

The equation (\ref{tise}) is equivalent to
\begin{equation}\label{equat1}
-\frac{\hbar^2}{2m}\frac{d^2\psi(a)}{da^2} + \left(
\frac{\hbar^2}{2m}\frac{3}{4}\frac{1}{a^2} -G M m\frac{1}{a} 
- \frac{\Lambda}{6}ma^2\right) \psi(a) = E\psi(a)\,.
\end{equation}
Let's solve the equation (\ref{equat1}) that is equivalent by dividing by $\frac{\hbar^2}{2m}$ to 
\begin{equation}
-\frac{d^2\psi(a)}{da^2} + \frac{2m}{\hbar^2}\left(
\frac{\hbar^2}{2m}\frac{3}{4}\frac{1}{a^2} -G M m\frac{1}{a} 
- \frac{\Lambda}{6}ma^2\right) \psi(a) = \frac{2m}{\hbar^2}E\psi(a)\,,
\end{equation}
labelling the parameters as follows
\begin{equation}
\alpha = \frac{3}{4};\:\: \delta = -\frac{4GMm^2}{\hbar^2};\:\:
\omega^2  = (\frac{-\Lambda}{3})\frac{m^2}{\hbar^2};\:\: k^2 = \frac{2m E}{\hbar^2}\,,
\end{equation}
we have 
\begin{equation}\label{equat2}
-\frac{d^2\psi(a)}{da^2} + \left(
\frac{\alpha}{a^2} + \frac{1}{2}\frac{\delta}{a} + \omega^2 a^2 -k^2
\right) \psi(a) = 0.
\end{equation}
The sign of the cosmological constant is for the moment left arbitrary.
In search of solution of the equation (\ref{equat2}), we set the Antsatz
\begin{equation}
\psi(a) = a^{\beta + 1}e^{-\frac{\omega}{2}a^2}v(a)\,,
\end{equation}
where $\beta(\beta + 1) = \alpha$ and the function $v(a)$ is an auxiliary function satisfying the equation 
\begin{equation}\label{aeq1}
v''(a)+ \left( 2(\beta +1)\frac{1}{a} - 2\omega a\right)v' (a)
+ \left( k^2 -\omega(2\beta +3) -\frac{1}{2}\frac{\delta}{a}\right) v (a) = 0\,.
\end{equation}
Performing the change of variable $x = w^{1/2}a$ in (\ref{aeq1}), we have 
\begin{equation}\label{equat3}
v''(x) + \left(\frac{2(\beta +1)}{x} -2x \right)v'(x) + 
\left( \frac{k^2}{\omega} -(2\beta +3) -\frac{1}{2}\frac{\delta \omega^{-1/2}}{x}\right) v(x) = 0.
\end{equation}
The equation (\ref{equat3}) is a particular case of the biconfluent Heun equation \cite{ronv} and its solution is the biconfluent Heun function
\begin{equation}
v(x) = \rm{HeunB}\left( 2\beta +1;\; 0;\; k^2\omega^{-1};\; \delta\omega^{-1/2};\;x\right)\,.
\end{equation}
Some details about biconfluent Heun equation are given in the Appendix (see after section (\ref{sec6})).

From the relation $\beta (\beta +1) = \alpha $ and $\alpha = 3/4$, 
there are two possibilities $\beta = 1/2$ or $\beta  = -3/4$.
For $(2\beta + 1)$ not a negative integer, the biconfluent Heun functions $v(x)$ can be written in series form as
\begin{eqnarray}
v(x) & =& \rm{HeunB}\left( 2\beta +1;\; 0;\; k^2\omega^{-1};\; \delta\omega^{-1/2};\;x\right);\\
&=& \sum_{p =0}^\infty\frac{A_p}{(1 + (2\beta +1))_p}\frac{x^p}{p!}\,,
\end{eqnarray}
where 
\begin{equation}
A_{p+2} = \frac{1}{2}\delta\omega^{-1/2}A_{p+1} -
(p+1)(p +1 + (2\beta +1))(k^2\omega^{-1} - (2\beta + 1) -2- 2p)A_p\,. 
\end{equation}
We consider the value $\beta = 1/2$ that gives $2\beta + 1 = 2$  and then 
\begin{eqnarray}
v(x) & =& \rm{HeunB}\left( 2;\; 0;\; k^2\omega^{-1};\; \delta\omega^{-1/2};\;x\right);\\
&=& \sum_{p = 0}^\infty\frac{A_p}{(3)_p}\frac{x^p}{p!},
\end{eqnarray}
where 
\begin{equation}\label{receq}
A_{p+2} = \frac{1}{2}\delta\omega^{-1/2}A_{p+1} -
(p+1)(p + 3)(k^2\omega^{-1} - 4 - 2p)A_p.
\end{equation}
From the recursion equation (\ref{receq}), the function $v(x)$ becomes a polynomial of degree n and the series terminate if and only if the two following conditions are fulfilled
\begin{enumerate}
\item $k^2\omega^{-1} - 4 = 2n $;
\item  $A_{n +1}$ = 0,
\end{enumerate}
where $A_{n+1}$ is a polynomial of degree $(n +1)$ in $\tilde{\delta} = \delta \omega^{-1/2}$. There are at most $(n+1)$ suitable values of value $ \tilde\delta $ usually labelled as $\tilde{\delta}_\mu^n, \: 0\le \mu \le n$. In that sense, we have 
\begin{eqnarray}
 v_n(x) & =& \rm{HeunB}\left( 2;\; 0;\; 2(n+2);\; \tilde{\delta}^n_\mu;\;x\right)\,;\\
 &=& P_{n,\mu}(2,0,x) = \sum_{p =0}^n\frac{A_p}{(3)_p}\frac{x^p}{p!}\,,
\end{eqnarray}
with
\begin{equation}
A_0 = 1\,;\; A_1 = \frac{1}{2}\tilde{\delta}\,;
\end{equation}
\begin{equation}
A_{p+2} = \frac{1}{2}\tilde{\delta}A_{p+1} -2
(p+1)(p + 3)(n-p)A_p\,;\quad A_{n+1} = 0.
\end{equation}
The eigenvalues are determined from the condition $k^2\omega^{-1} -4 = 2n$ as follows
\begin{equation}
E_n = (n +2)\hbar \left(\frac{-\Lambda}{3}\right)^{\frac{1}{2}}\,.
\end{equation}
The associated physically acceptable wave functions $\psi_n(a)$ are 
given by  
\begin{equation}\label{wavfunc}
\psi_n(a) = C_n a^{3/2}e^{-\frac{\omega}{2}a^2}\rm{HeunB}
\left( 2 ;\; 0;\;2(n + 2) ;\tilde\delta_\mu^n;\;\omega^{\frac{1}{2}}a\right)\,.
\end{equation}
The constant $C_n$ are arbitrary factors that can be determined under the normalization condition
\begin{equation}
\int_0^\infty \vert \psi_n(a)\vert ^2 da = 1\,.
\end{equation}
We have 
\begin{equation}
C_n^2 \int_0^\infty a^3 e^{-\omega a^2} \sum_{p =0}^n\frac{A_p}{(3)_p}\frac{\omega^{\frac{p}{2}}a^p}{p!}\sum_{s =0}^m\frac{A_s}{(3)_s}\frac{\omega^{\frac{s}{2}}a^s}{s!}da = 1\,,
\end{equation}
\begin{equation}
C_n^2\sum_{p=0}^n\sum_{s =0}^m\frac{A_p\omega^{\frac{p}{2}}}{(3)_p p!}\frac{A_s\omega^{\frac{s}{2}}}{(3)_s s!}
\int_0^\infty a^{3+p+s}e^{-\omega a^2} da = 1\,.
\end{equation}
For $ p = s$, $\mu = \nu$ and $m=n$, we have
\begin{equation}\label{formu1}
C_n^2 \sum_{p= 0}^n\frac{A_p^2\omega^p}{[(3)_p p!]^2}\int_0^\infty a^{3+2p}e^{-\omega a^2}da = 1.
\end{equation}
Let's use now the formula 
\begin{equation}
\int_0^\infty y^{\lambda -1} e^{-\eta y^u}dy = \frac{1}{u}\eta^{-\lambda/u}\Gamma(\frac{\lambda}{u})\,,
\end{equation}
so we have 
\begin{equation}\label{formu2}
\int_0^\infty a^{3+2p}e^{-\omega a^2}da = \frac{1}{2}\omega^{-(2 +p)}\Gamma(2 +p)\,,
\end{equation}
and inserting (\ref{formu2}) into (\ref{formu1}), we have
\begin{equation}\label{formu3}
C_n^2 \sum_{p= 0}^n\frac{A_p^2\omega^p}{[(3)_p p!]^2}
\frac{1}{2}\omega^{-(2 +p)}\Gamma(2 +p) = 1\,,
\end{equation}
that is 
\begin{equation}
C_n^2\sum_{p = 0}^n \frac{A_p^2}{ [(3)_p p!]^2 }\frac{1}{2\omega^{2}}\Gamma(2 +p) = 1\,,
\end{equation}
so 
\begin{equation}\label{intconst}
C_n =\left[ \sum_{p = 0}^n \frac{A_p^2}{ [(3)_p p!]^2 }\frac{1}{2\omega^{2}}\Gamma(2 +p)\right]^{-\frac{1}{2}}\,.
\end{equation}
Inserting the constant of integration $C_n$ in equation (\ref{intconst}) into the equation (\ref{wavfunc}), we have 
\begin{eqnarray}\nonumber
\psi_n(a) &=& \left[ \sum_{p = 0}^n \frac{A_p^2}{ [(3)_p p!]^2 }\frac{1}{2\omega^{2}}\Gamma(2 +p)\right]^{-\frac{1}{2}}\\
&\times & a^{\frac{3}{2}}e^{-\frac{\omega}{2}a^2}\rm{HeunB}
\left( 2 ;\; 0;\;2(n +2) ;\tilde\delta_\mu^n;\;\omega^{\frac{1}{2}}a\right)\,.
\end{eqnarray}

We consider now the case of matter dominated by a negative cosmological constant, so $\Lambda  =  - |\Lambda|$. The  wave functions for  the Newtonian universe with negative cosmological constant are given by
\begin{eqnarray}\nonumber
\psi_n(a) = \left[ \sum_{p = 0}^n \frac{A_p^2}{ [(3)_p p!]^2 }\frac{1}{2\omega^{2}}\Gamma(2 +p)\right]^{-\frac{1}{2}}\\\label{leq}
a^{\frac{3}{2}}e^{-\frac{\omega}{2}a^2}\rm{HeunB}
\left( 2 ;\; 0;\;2(n +2) ;\tilde\delta_\mu^n;\;\omega^{\frac{1}{2}}a\right)\,;
\end{eqnarray}
\begin{equation}
E_n = (n +2)\hbar \left(\frac{|\Lambda|}{3}\right)^{\frac{1}{2}},\: n= 0, 1, 2 \ldots;
\end{equation}
with the parameters given by
\begin{equation}
\omega^2 = \frac{|\Lambda|}{3}\frac{m^2}{\hbar^2};\quad
\omega = \left(\frac{|\Lambda|}{3}\right)^{\frac{1}{2}}\frac{m}{\hbar};\quad
\omega^{\frac{1}{2}} = \left( \frac{|\Lambda|}{3}\right)^{\frac{1}{4}}\left( \frac{m}{\hbar}\right)^{\frac{1}{2}}\,;
\end{equation}
\begin{equation}
\tilde{\delta}= -4GM \left( \frac{|\Lambda|}{3}\right)^{-\frac{1}{4}}\left( \frac{m}{\hbar}\right)^{\frac{3}{2}}\,.
\end{equation}

\section{Concluding remarks}\label{sec5}

We consider a small spherical portion of the universe, where Newton's theory applies and the behavior of this portion may reflect the evolution of the universe.
Since the variable $a(t)$ that is referred to as the scale factor is positive, we choose to pay attention to the affine quantization procedures. The solution of the time independent Schr\"odinger equation is of type biconfluent Heun function where the two first parameters are fixed to $2$ and $0$ respectively.  Our results are quite similar to the ones in \cite{bras2} with the difference that performing affine quantization, the Hamiltonian operator gains an extra term that is proportial to $1/a^2$ and that extra term fixes the first parameter of the biconfluent Heun function to the value of $2$. In a case of matter dominated by negative cosmological constant, $\Lambda = - |\Lambda|$, the eigenvalues are positive equally spaced  and non-degenerated.

The Newtonian approach is much simple from the conceptual and mathematical point of views however it met also problems and criticisms \cite{prob1, prob2, prob3, prob4, prob5}. It is interesting to work out under which conditions Newtonian cosmology applies. Recent works in Newtonian cosmology include \cite{barrow, casadio}.

\section*{Appendix: Biconfluent Heun Equation}\label{sec6}

A biconfluent Heun equation, denoted by 
$\rm{BHE}(\alpha,\beta,\gamma,\delta)$ is the equation of form
\begin{equation}
xu''(x) + \left[ 1 +\alpha -\beta x -2 x^2 \right]u'(x) 
+ \left\lbrace 
\left( \gamma -\alpha -2 \right) x -
\frac{1}{2}\left(\delta + (\alpha +2)\beta \right)
\right\rbrace u(x) = 0
\end{equation}
in which $(\alpha, \beta,\gamma,\delta) \in \mathbb{C}^4$. It has a regular singular point at $0$ and an irregular singular point at 
$\infty$. The biconfluent Heun equation is well known and studied in a mathematical point of view (\cite{ronv}, \cite{math1, math2, math3, math4} ) and has some applications in different areas of physics \cite{phys1, phys2, phys3}. Refering to [ \cite{ronv}, pp. 203 - 206], if the biconfluent Heun equation admits a polynomial solution then it is necessary that $\gamma -\alpha -2p = 2n$, where $n$ is some non-negative integer, holds. When $\alpha$ is not a negative integer, one can denote by $N(\alpha,\beta, \delta, x)$ a power series (analytic) solution that can be written as 
\begin{equation}
N(\alpha,\beta,\gamma, \delta, x) = \sum_{p =0}^\infty \frac{A_p}{(1 +\alpha)_p}\frac{x^p}{p!},
\end{equation}
with $(\alpha)_p = \frac{\Gamma(\alpha + p)}{\Gamma(\alpha)},\; p\ge 0$, and satisfies the three -term recursion formula 
\begin{eqnarray}\nonumber
A_{p+2} & =& \left\lbrace
(p + 1)\beta + \frac{1}{2}[\delta + \beta (1 +\alpha)]
\right\rbrace A_{p+ 1} \\ \label{eqrec}
&-& (p +1)(p+1+\alpha)(\gamma -\alpha -2-2p)A_p = 0\,,
\end{eqnarray}
where $A_0 = 1,\: A_1 = \frac{1}{2}[\delta + \beta (1 + \alpha)]$. The three recursion term in (\ref{eqrec}) terminates if and only if 
$\gamma -\alpha -2 = 2n,$ and  $ A_{n +1} = 0$ simultaneously where $n$ is some non negative integer. Performing induction $A_{n+1}(\delta)$ is a polynomial of $\delta$ of degree $n +1$ hence possessing at most $n+1$ roots $\delta_\mu^n, \mu = 0,1,2,\ldots$ When the series solution terminates, then we write
\begin{equation}
P_{n,\mu}(\alpha,\beta; x) = N(\alpha, \beta, \alpha + 2(n +1),\delta_\mu^n;x), \quad 0 \le \mu \le n,\: n = 0,1,2,\ldots
\end{equation}
When $\alpha +1 > 0$ and $\beta \in \mathbb{R}$, then the $(n+1) $ 
roots are real. When all the roots are simple, then the polynomial solutions described are precisely orthogonal polynomials (see \cite{ronv}, pargraph 3.3).

\section*{Acknowledments}

I am grateful to Sana Khadim PhD student at NUST-Islamabad/Pakistan whose interest in quantization of black holes inspired and motivated me to include quantum cosmology in my research interests.

\end{document}